\documentclass[10pt,
							 twocolumn,
							 superscriptaddress,
							 english,
							 prl,
							 showpacs,
							 floatfix,
							 aps
							]{revtex4-1}
\usepackage[utf8]{inputenc}
\usepackage{amsmath}
\usepackage{amssymb}
\usepackage{graphicx}
\usepackage{xspace}

\makeatletter
\@ifundefined{textcolor}{}
{%
 \definecolor{BLACK}{gray}{0}
 \definecolor{WHITE}{gray}{1}
 \definecolor{RED}{rgb}{1,0,0}
 \definecolor{GREEN}{rgb}{0,1,0}
 \definecolor{BLUE}{rgb}{0,0,1}
 \definecolor{CYAN}{cmyk}{1,0,0,0}
 \definecolor{MAGENTA}{cmyk}{0,1,0,0}
 \definecolor{YELLOW}{cmyk}{0,0,1,0}
}

\usepackage{soul}
\usepackage{braket}
\usepackage[backref=none,
bookmarksnumbered=true,
bookmarks=true,
bookmarksopen=true,
colorlinks=true,
citecolor=blue,
linkcolor=blue,
anchorcolor=green,
urlcolor=blue,unicode=false]{hyperref}

\renewcommand{\v}[1]{\ensuremath{\mathbf{#1}}} 
\newcommand{\gv}[1]{\ensuremath{\mbox{\boldmath$ #1 $}}} 
 
\let\baraccent=\= 
\renewcommand{\=}[1]{\stackrel{#1}{=}} 

\newcommand{\unitspace}{~}

\newcommand{\didv}{\ensuremath{\mathrm{d}I/\mathrm{d}V}\xspace}

\newcommand{\DeltaT}{\ensuremath{\Delta_\mathrm{tip}}\xspace}

\newcommand{\Fig}[1]{Fig.\unitspace\ref{fig:#1}}
\newcommand{\Figure}[1]{Figure\unitspace\ref{fig:#1}}

\newcommand{\YSR}{YSR\xspace}
\newcommand{\udir}[1]{$\left<#1\right>$}

\newcommand{\dorb}[1]{$d_\mathrm{#1}$}

\newcommand{\Mn}[2]{$\mathrm{Mn}^\mathrm{#2}_\mathrm{Pb(#1)}$\xspace}

\DeclareMathOperator{\unm}{\unitspace\mathrm{nm}}

\DeclareMathOperator{\umV}{\unitspace\mathrm{mV}}
\DeclareMathOperator{\umuV}{\unitspace\mathrm{\mu V}}
\DeclareMathOperator{\ueV}{\unitspace\mathrm{eV}}

\DeclareMathOperator{\umueV}{\unitspace\mathrm{\mu eV}}

\DeclareMathOperator{\upA}{\unitspace\mathrm{pA}}

\DeclareMathOperator{\uK}{\unitspace\mathrm{K}}

\DeclareMathOperator{\uHz}{\unitspace\mathrm{Hz}}
\DeclareMathOperator{\umin}{\unitspace\mathrm{min}}
\DeclareMathOperator{\umbar}{\unitspace\mathrm{mbar}}
\DeclareMathOperator{\uAA}{\unitspace\mathrm{\AA{}}}

\makeatother

\usepackage{babel}

\begin{document}
\title{Orbital Picture of Yu-Shiba-Rusinov Multiplets}

\author{Michael Ruby}
\affiliation{\mbox{Fachbereich Physik, Freie Universit\"at Berlin, 14195 Berlin, Germany}}

\author{Yang Peng}
\affiliation{\mbox{Dahlem Center for Complex Quantum Systems and Fachbereich Physik, Freie Universit\"at Berlin, 14195 Berlin, Germany}}

\author{Felix von Oppen}
\affiliation{\mbox{Dahlem Center for Complex Quantum Systems and Fachbereich Physik, Freie Universit\"at Berlin, 14195 Berlin, Germany}}

\author{Benjamin W. Heinrich}
\affiliation{\mbox{Fachbereich Physik, Freie Universit\"at Berlin, 14195 Berlin, Germany}}

\author{Katharina J. Franke}
\affiliation{\mbox{Fachbereich Physik, Freie Universit\"at Berlin, 14195 Berlin, Germany}}
\date{\today}
\begin{abstract}
We investigate the nature of Yu-Shiba-Rusinov (\YSR) subgap states induced by single manganese (Mn) atoms adsorbed on different surface orientations of superconducting lead (Pb). 
Depending on the adsorption site, we detect a distinct number and characteristic patterns of \YSR states around the Mn atoms. 
We suggest that the \YSR states inherit their properties from the Mn $d$-levels, which are split by the surrounding crystal field. 
The periodicity of the long-range \YSR oscillations allows us to identify a dominant coupling of the $d$-states to the outer Fermi sheet of the two-band superconductor Pb. %
\end{abstract}
\pacs{%
			} 
\maketitle 


Local magnetic moments in metals induce potential and exchange scattering of quasiparticles. When the metal enters the superconducting state, this leads to the formation of localized bound states within the superconducting gap, referred to as Yu-Shiba-Rusinov (\YSR) states \cite{Yu,Shiba,Rusinov69}. In the simplest picture, the magnetic moment is viewed as a classical impurity spin, which is exchange coupled to itinerant electrons with an isotropic Fermi surface. Treating the exchange coupling as local and isotropic, a single particle-hole symmetric pair of \YSR states is predicted within the gap whose wave functions oscillate with a wavelength $\lambda_F$ ($\lambda_F$ being the Fermi wavelength) \cite{Fetter65,Rusinov69,Yazdani97,reviewRMP}. Anisotropy of the Fermi surface induces a scattering pattern reflecting the symmetry of the host lattice, as discussed theoretically and observed in recent experiments \cite{Salkola97,Menard15}. 

With sufficient resolution, experiments show not only one, but several pairs of \YSR resonances \cite{Ji08, Ruby15, Hatter15}. The origin of multiple \YSR resonances was assigned to scattering channels with different angular momenta ($l=0,1,2,\ldots$) \cite{Flatte97, Ji08}, or to the anisotropy splitting of the magnetic states of the adsorbate \cite{Zitko11, Hatter15}. The arguments were based solely on the energetic alignment of the \YSR states and not on the spatial extension and patterns of the states, which would allow one to establish a link with the orbital structure of the magnetic impurity.

Here, we address the origin of multiple \YSR states by combining scanning tunneling microscopy and spectroscopy (STM and STS) experiments, which are powerful tools to map the energetic and spatial characteristics of energy levels, with a theoretical analysis. 

Our study is based on Mn adatoms placed on a Pb substrate. The main advantage of this system is that the Mn adatoms are expected to be in the Mn$^{++}$ configuration with five $d$-electrons. According to Hund's rules, the Mn $d$-shell is in a $^6$S$_{5/2}$ configuration and, hence, spherically symmetric. Thus, the ion cannot change the angular momentum of the conduction electrons in an isotropic environment and in the absence of spin-orbit coupling, which facilitates comparison between experiment and theory. The $s$-wave superconductor Pb is an experimentally well-studied substrate due to its high critical temperature ($T_{\mathrm{c}} = 7.2\uK$), which can be readily prepared by standard ultrahigh vacuum preparation techniques \cite{Ji08, Ruby15,Hatter15, Ruby14}. It also constitutes an appealing substrate for topologically nontrivial nanostructures \cite{yazdani, pawlak2015afm, RubyMaj15, roetynen15, Kim15, menard2016, li16}. Therefore, it would be rewarding to develop a more systematic understanding of magnetic adatom systems, reaching all the way from monomers and dimers to chains or even two-dimensional arrays. 
Because Pb possesses two disjunct Fermi surfaces, it is a two-band superconductor with two distinct gaps~\cite{Floris07}, which can be resolved in STM experiments~\cite{Ruby14}. Our study provides evidence that the \YSR states in this system are predominantly associated with one of the two bands.

The experiments were carried out in a {\sc Specs} JT-STM under ultrahigh vacuum conditions at a temperature of $1.2\uK$. The Pb single crystals were cleaned by Ne$^+$ ion sputtering ($900\ueV$, $1.5\times10^{-4}\umbar$, background pressure $<1.5\times10^{-9}\umbar$). Annealing to $430\uK$ for $30\umin$ results in clean, flat, and superconducting terraces. Spectra of the differential conductance \didv as a function of sample bias $V$ were acquired with a standard lock-in technique at a frequency of $912\uHz$. To achieve high energy resolution, we cover etched W-tips with Pb by deep indentations into the clean Pb surface until superconductor-superconductor tunneling spectra are measured. The use of a superconducting tip together with an elaborate grounding and RF-filtering setup yields effective energy resolutions of $\approx60\umueV$ at $1.2\uK$. 
Using superconducting tips involves a convolution of the densities of states of tip and substrate, so that all subgap states $\epsilon$ appear shifted by the superconducting gap of the tip ($\DeltaT$) to an energy $eV=\pm (\epsilon+\DeltaT)$~\cite{note}.
Mn adatoms were evaporated onto the clean sample in the STM at a temperature below $15\uK$, resulting in a density of $\simeq100$ atoms per $100\times100\unm^2$.

\begin{figure}[tb]
	\includegraphics[width=\columnwidth]{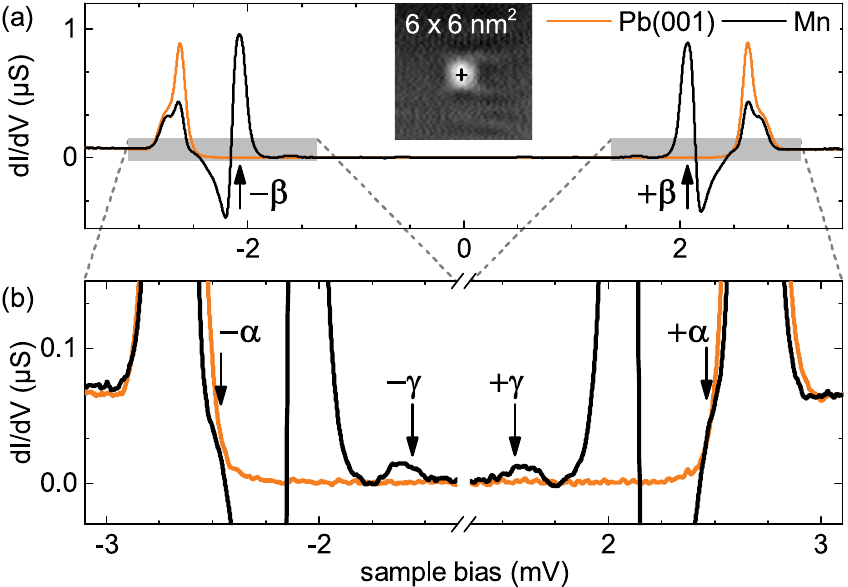}
	\caption{ (a,b) \didv spectrum of a Mn adatom ({\it black}) and of clean Pb(001) ({\it orange}). The inset shows a topography of the adatom. Three subgap resonances $\pm\alpha$, $\pm\beta$, and $\pm\gamma$ are marked by arrows. 
 Set point: $300\upA, 5\umV$; modulation: $15\umuV_\mathrm{rms}$.}
	\label{fig:100a}
\end{figure}

We first deposit Mn atoms on the Pb(001) surface. All Mn adatoms adsorb in equivalent sites, which we call \Mn{001}{}. The adsorption site is stable against manipulation with the tip. 
In topography, the adatoms display a fourfold shape and a height of $0.15\uAA$ at $5\umV$ [\Fig{100a}(a), inset]. 
Spectra on top of the adatoms reveal three pairs of \YSR resonances inside the superconducting gap. The one with the largest spectral intensity (labeled $\pm\beta$) is found at a bias voltage of $\simeq\pm2.08\umV$ [\Fig{100a}(a)], complemented by two faint resonances at $\simeq\pm2.47\umV$ and $\simeq\pm1.61\umV$ [\Fig{100a}(b)]. We label the latter resonances as $\pm\alpha$, and $\pm\gamma$, respectively. 

\begin{figure}[tb]
	\includegraphics[width=\columnwidth]{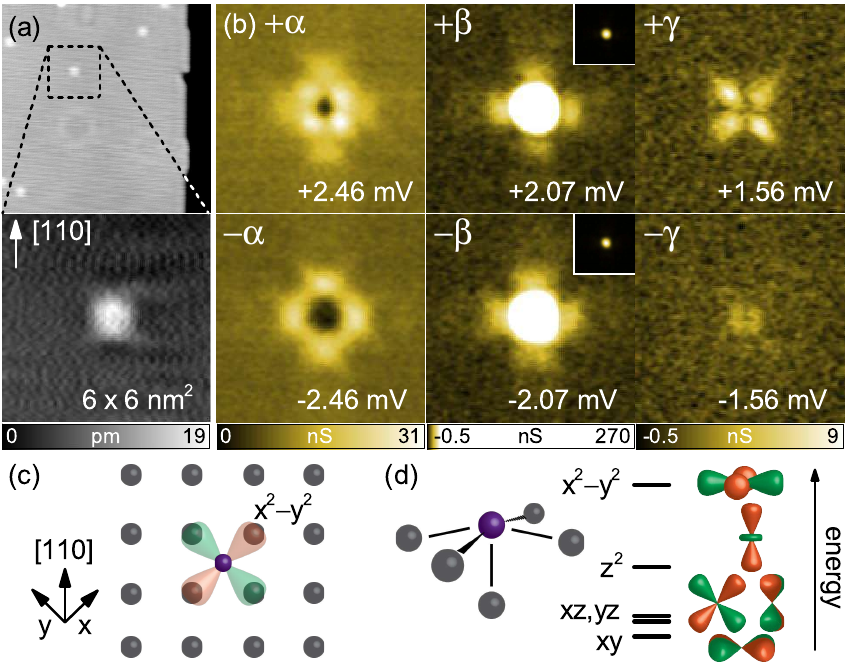}
	\caption{(a) Topography of a Mn adatom on Pb(001). Set point: $50\umV$ (zoom: $5\umV$), $150\upA$. (b) \didv maps  of the adatom depicted in (a) at the energy of the subgap resonances $\pm\alpha$, $\pm\beta$, and $\pm\gamma$ (feedback opened in each pixel at $5\umV$ and $150\upA$; modulation: $25\umuV_\mathrm{rms}$). 
The color scale of $\pm\beta$ is stretched  to enhance low intensity features (inset shows $\pm\beta$ with linear scale). 
The center is one hundred times more intense than the fourfold symmetric lobes.
 (c) Schematic top-view of the adsorption of a Mn adatom in the (001) hollow site.  
(d) Corresponding crystal field splitting of the $d$-levels.}
	\label{fig:100b}
\end{figure}

The spatial patterns of all three \YSR states show characteristic fourfold symmetries [\Fig{100b}(b)], with extensions up to $\approx1.6\unm$. This is an order of magnitude larger than the atomic radius. The resonances $\pm\beta$ show intensity mainly at the center of the impurity, with some weak intensity along the \udir{110} directions. 
Because of their large intensity we observe a negative differential conductance (NDC) [see \Fig{100a}(a,b)]. The maps at $\pm\alpha$ are dominated at the center by this NDC because of the energetic overlap of $\pm\beta$ with $\pm\alpha$; hence, the maps show no spectral intensity here. The intensity is largest at a distance of $\simeq0.9\unm$ away from the center.  The map at $+\gamma$ resolves a clover leaf pattern along the \udir{100} directions, whereas we hardly detect any signal for $-\gamma$.  

The \YSR patterns resemble the shape of $d$-orbitals and thus suggest a correlation of the \YSR resonances and the orbitals hosting an unpaired electron spin. This requires a splitting of the $d$-states of Mn due to the crystal field imposed by the adsorption site. In a hollow site, the nearest neighbors form a square pyramidal coordination symmetry, which removes the degeneracy of the five $d$-levels [\Fig{100b}(c,d)]. According to simple arguments of crystal field theory, the \dorb{x^2-y^2}-orbital lies highest, followed by the \dorb{z^2}-orbital, the degenerate \dorb{xz}- and \dorb{yz}-orbitals, and the \dorb{xy}-orbital at the lowest energy.  The energy separation between \dorb{xz/yz} and \dorb{xy} depends on the ratio of in-plane and out-of-plane bonding distances and the levels become degenerate for an adsorption configuration with all distances being equal. Indeed, we find hints that resonance $\pm\gamma$ is composed of almost degenerate states, as it splits up upon interaction with neighboring atoms (see Supplemental Material~\cite{Supplementary}). 

Simple models of \YSR states rely on scattering of $l=0$ conduction electrons (for a notable exception, see Ref.~\cite{Moca08}). However, as emphasized by \mbox{Schrieffer} \cite{Schrieffer67}, only $l=2$ conduction electrons are (potential and exchange) scattered by Mn$^{++}$ impurities in an isotropic metal, which is a consequence of their $S$-state nature. Starting with the isotropic case, we can then account for lattice and surface effects by the addition of anisotropic crystal fields which (partially) remove the degeneracy between the $d$-levels and make the potential and exchange coupling with the impurity orbital dependent~\cite{Moca08} (as follows from a standard Schrieffer-Wolff transformation \cite{Schrieffer67,Hewson}). This structure is then inherited by the \YSR states (see Supplemental Material~\cite{Supplementary}). This picture suggests that Mn$^{++}$ impurities actually induce five pairs of \YSR states whose degeneracies and spatial patterns reflect the crystal-field-split $d$-orbitals.

\begin{figure*}[t]
	\includegraphics[width=\textwidth]{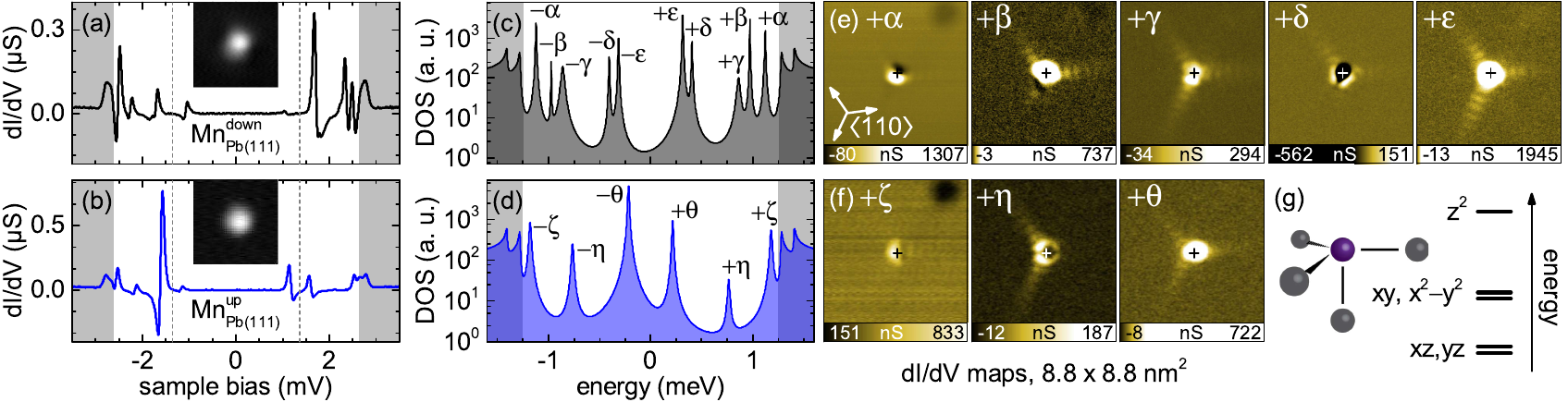}
	\caption{(a,b) \didv spectra of the same Mn adatom on Pb(111) in the two adsorption sites as indicated in the figure. The shaded areas mark the two BCS coherence peaks \cite{Ruby14} and the normal state. The dashed lines indicate the tip gap ($\pm1.38\umV$). 
Set point: $100\upA, 5\umV$; lock-in modulation: $15\umuV_\mathrm{rms}$. The insets show topographies of the adatom in the respective adsorption site ($2.6\times2.6\unm^2$). (c) Deconvolved sample density of states of   \Mn{111}{down} exhibits five \YSR resonances ($\pm\alpha$,\ldots,$\pm\varepsilon$). (d) Deconvolved density of states of \Mn{111}{up} shows three \YSR resonances ($\pm\zeta$,$\pm\eta$,$\pm\theta$). 
(e,f) \didv maps of the \YSR resonances in both adsorption sites  (feedback opened in each pixel at $5\umV$ and $400\upA$; modulation: $20\umuV_\mathrm{rms}$). Crosses denote the same position in all maps. The dark spot in the top right corner is a sub-surface neon inclusion. 
The color scale is stretched to give the best contrast to spatially extended features (for maps with linear color scale see Supplemental Material~\cite{Supplementary}). (g) Crystal field splitting of the $d$-levels for an adatom in a hollow site.}
	\label{fig:111}
\end{figure*}

We now aim for an identification of the specific $d$-orbitals that give rise to the \YSR resonances $\pm\alpha$, $\pm\beta$, and $\pm\gamma$. 
The fact that $\beta$ is the most intense resonance indicates that it has the largest wave function overlap with the tip [note the color scale in \Fig{100b}(b)]. Moreover, the main intensity is spherically symmetric [inset of \Fig{100b}(b)]. Both arguments suggest that $\beta$ originates from scattering at the \dorb{z^2}-orbital, which is oriented along the surface normal. 
We note that both resonance $\alpha$ and $\gamma$ exhibit the largest intensity along the \udir{100} directions, i.e., towards the nearest neighbors. Hence, an assignment solely based on directions is not a priori possible. Instead, we rely on the above mentioned observation of the degeneracy of $\pm\gamma$. It includes the \dorb{xz,yz}- and \dorb{xy}-orbitals as scattering centers, whereby the latter is oriented in plane and thus only contributes weakly to tunneling. Then, resonance $\alpha$ is induced by scattering from the \dorb{x^2-y^2}-orbital.

In addition to the influence of the orbital symmetry, the long-range scattering pattern obtains structure from the anisotropy of the Fermi surface \cite{Salkola97}. In case of Pb(001), the projected Fermi surface obeys a $C_\mathrm{4}$ symmetry. Electron (hole) propagation along the \udir{110} directions appears enhanced due to focusing perpendicular to the low-curvature regions of the Fermi surface \cite{Kurnosikov09, Weismann09, Ruby14}. The anisotropy of the Fermi surface thus amplifies the $C_\mathrm{4v}$ angular dependence of the $d$-orbitals. This imprints a faint fourfold shape on the \YSR patterns. In the case of the \dorb{z^2}-orbital ($\pm\beta$) it is one hundred times smaller than the spherically symmetric central part of the resonance [compare \Fig{100b}(b) with insets].


In order to test the validity of our model, we carried out similar experiments on Pb(111). This surface imposes a different crystal field on the adsorbate so that we expect different characteristic \YSR energies and patterns. Deposition of Mn on Pb(111) leads to a unique adsorption site for all adatoms. In topography, they appear with a height of $\approx0.5\uAA$ at $50\umV$ and a slightly oval shape along one of the three \udir{110} directions [see inset in \Fig{111}(a)]. 

By approaching the STM tip on top of a Mn adatom at $V= +5\,$mV until contact formation, the atom is transferred from the initial adsorption site to a site with a larger apparent height ($\approx1.1\uAA$ at $50\umV$) and a fully symmetric appearance in topography~\cite{Ruby15}.
The initial adsorption configuration is recovered by contact formation at $V=-180$\,mV, which yields the original height and shape, with the oval shape being oriented along one of the three \udir{110} directions, though not necessarily the initial one. We refer to the two adsorption sites according to their apparent heights as \Mn{111}{down} and \Mn{111}{up}, respectively. 

Both adsorption sites show several \YSR resonances inside the superconducting energy gap at $eV=\pm(\epsilon+\DeltaT)$ [\Fig{111}(a,b)]. [In addition, we also observe resonances at $eV=\pm(\DeltaT-\epsilon)$ which originate from thermally activated tunneling into or out of \YSR states~\cite{Ruby15} and are restricted to small $\epsilon$ at $1.2\uK$.] The deconvolved density of states is plotted in \Fig{111}(c,d) \cite{RubySF16}. Interestingly, we observe different numbers of \YSR resonances for the two adsoption sites, in addition to shifts in energy. For  \Mn{111}{down} adatoms, we resolve five \YSR resonances, independent of the direction of the oval appearance. In contrast, \Mn{111}{up} adatoms exhibit only three resonances.

The multiplicity of the \YSR states is consistent with certain adsorption sites. The threefold multiplicity of the \Mn{111}{up} adsorption site agrees with a hollow site, which is subject to a trigonal pyramidal crystal field. This induces a $d$-level splitting with the \dorb{z^2}-orbital lying highest in energy, followed by the degenerate \dorb{xy}- and \dorb{x^2-y^2}-orbitals. Lowest in energy are the degenerate \dorb{xz}- and \dorb{yz}-orbitals [\Fig{111}(g)]. 
The fivefold multiplicity of the \YSR resonances of \Mn{111}{down} indicates the removal of all degeneracies of the $d$-orbitals. This is the case when the atom is slightly displaced from a hollow site, which is consistent with its oval-shaped appearance.

Next, we investigate the spatial distribution of the \YSR resonances by \didv maps at the respective energies. They are shown for positive bias voltages in \Fig{111}(e) for \Mn{111}{down} and in \Fig{111}(f) for the same atom after manipulation into the \Mn{111}{up} adsorption state. 
Maps at negative bias voltages reveal similar patterns (see Supplemental Material~\cite{Supplementary}).
The maps do not reflect the typical fourfold shape of the Mn $d$-orbitals. The $C_\mathrm{3v}$ symmetry of the ligand field polarizes the $d$-orbitals due to hybridization with the $p$-orbitals \cite{JeanBook}. As a result, the characteristic $d$-orbital shapes are deformed, resulting in an overall twofold symmetry as reflected in the \YSR maps. We may tentatively assign the \YSR states by arguments of wave function overlap with the tip.
Both, $\pm\theta$ and $\pm\zeta$ show a large intensity signifying an out-of-plane extension of the wave function. 
The spherical symmetry of $\pm\theta$ at the impurity site suggests that it originates from scattering at the \dorb{z^2}-orbital. Resonances $\pm\zeta$ would thus correspond to the degenerate \dorb{xz}- and \dorb{yz}-orbitals. The in-plane \dorb{x^2-y^2}- and \dorb{xy}-orbitals possess the smallest wave function overlap and hence the lowest intensity at $\pm\eta$. 


\begin{figure}[bt]
	\includegraphics[width=\columnwidth]{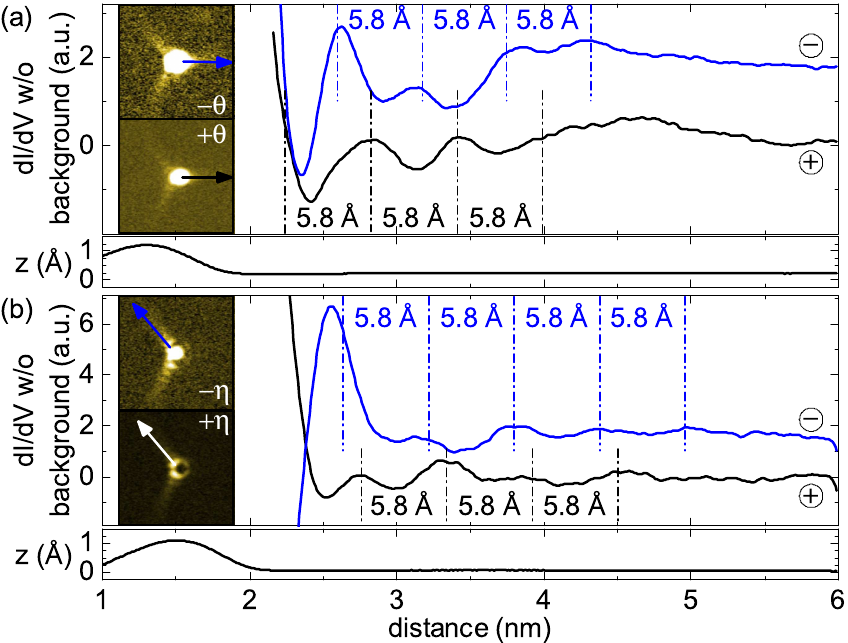}
	\caption{Lateral evolution of the spectral intensity at positive (black) and negative bias (blue) of the two \YSR resonances with lowest binding energy for \Mn{111}{up}. Set point: $400\upA, 4\umV$; modulation: $20\umuV$. The blue curve is offset by +1.5. The $1/r^2$ decay has been removed (for full data see Supplemental Material \cite{Supplementary}). The insets show the \didv maps at the corresponding energies. The arrows mark the direction along which the intensity is plotted. The $z$ profile gives the apparent height along the distance from the impurity center.}
	\label{fig:phase}
\end{figure}

Another interesting feature is the large lateral extension of several of the \YSR states. 
Three ($\beta$,$\gamma$,$\varepsilon$) of the five states of the \Mn{111}{down} adatoms and two ($\eta$, $\theta$) of the three states of \Mn{111}{up} persist up to $4\unm$ away from the adsorbate [\Fig{111}(e,f)]. The beamlike extension along the \udir{110} axes is due to the focusing of scattering electrons from the flat parts of the Fermi surface. We identify oscillating intensities within the beams. Indeed, an oscillation with $2\,k_\mathrm{F}$ is expected for \YSR states, because their wave functions obey \cite{Rusinov69}
\begin{equation}
\psi^\pm(r) \propto \frac{\sin{\left({k_\mathrm{F} r + \delta^\pm}\right)}}{k_\mathrm{F} r}~\exp{\left(-\left|\sin{\left(\delta^+ - \delta^-\right)}\right| \frac{r}{\xi}\right)} ~.
\label{eq:ShibaWF}
\end{equation}
Here, $k_\mathrm{F}$ is the Fermi wave vector, $\xi$ is the coherence length, and $\delta^\pm$ are the scattering phase shifts of the \YSR states at positive and negative bias, respectively.  
The electron density $\left|\psi(r)\right|^2$ thus decays as $1/r^2$. We have removed this dependence by a fit to the decaying intensity of states $\pm\theta$ and $\pm\eta$ (for details see Supplemental Material~\cite{Supplementary}). The result is plotted in \Fig{phase} and highlights the oscillations. We observe up to four periods with a periodicity of $\simeq5.8\uAA$, which should be compared to the Fermi wavelength of the substrate. Pb possesses two disjunct Fermi sheets \cite{Floris07}. One sheet is $s$-$p$-like and originates from the second Brillouin zone, the other one is $p$-$d$-like and originates from the third Brillouin zone. The peculiar band structure gives rise to two superconducting energy gaps \cite{Floris07, Ruby14}, which we also observe in the \didv spectra. The corresponding Fermi wavelengths along the \udir{110} direction are $\lambda_\mathrm{F}=7.8\pm0.8\uAA$ for the first, and $\lambda_\mathrm{F}=12.1\pm0.5\uAA$ for the second band \cite{Lykken71}. The observed periodicity agrees with $\lambda_\mathrm{F}/2$ of the second Fermi sheet. Hence, the \YSR resonances arise due to magnetic scattering with electrons in the $p$-$d$-like band. This is a reasonable conjecture in view of the more localized character of this band compared to the more delocalized nature of the $s$-$p$-like band. 

The oscillations of hole-like and electron-like \YSR resonances are phase shifted in \Fig{phase} with a larger shift between resonances  $\pm\theta$ compared to resonances  $\pm\eta$. 
This agrees with the dependence of the binding energies of the \YSR states on the phase shifts according to $\epsilon=\Delta \cos(\delta^+ - \delta^-)$. It implies that the closer states are to the gap edge the smaller is the phase shift between the positive and negative \YSR component. 

To summarize, we investigated \YSR states of transition metal adatoms on high symmetry surfaces of the BCS superconductor Pb. The adsorption site imposes a distinct crystal field splitting on the $d$-orbitals. 
We could show that the \YSR states inherit the symmetry of the scattering potential from the individual $d$-orbitals of the adatom. On the Pb(001) surface, spatially resolved conductance maps allow us to identify the corresponding $d$-orbitals. The strong influence of the anisotropic Fermi surface overwhelms this assignment on the Pb(111) surface. The oscillatory patterns reveal the Fermi wavelength of the $p$-$d$-like Fermi sheet to be responsible for the scattering pattern. The long-range and directional nature of the states are promising for the design of coupled adatom structures. 

 {\it Note added.}  In the final stage of preparing the manuscript, we became aware of related work on Cr atoms on Pb(111)~\cite{Choi16}.


We acknowledge funding by the Deutsche Forschungsgemeinschaft through Grant No. FR2726/4 and through collaborative research Grants No. Sfb 658, No. CRC 183, and No. SPP 1666, as well as by the European Research Council through Consolidator Grant NanoSpin.

\bibliographystyle{apsrev4-1}

\begin{thebibliography}{99}
\bibitem{note} Note that the spectra reflect the density of states of the \YSR states because they are recorded in the weak tunneling regime~\cite{Ruby15}.
\end{thebibliography}

\clearpage

\renewcommand{\theequation}{S\arabic{equation}}

\onecolumngrid

\section*{\Large{Supplementary Material}}

\vspace{0.7cm}

\section{Theoretical Background}

The Manganese (Mn) adatoms are presumably in a $^6$S$_{5/2}$ configuration. When placed in an isotropic environment, this implies that the exchange interaction is with the $l=2$ conduction electrons and conserves angular momentum \cite{Schrieffer67}. Lifting the degeneracy between the $d$-levels is the result of crystal-field splittings reflecting the anisotropy of the host, and the resulting multiplicities are largely determined by symmetry considerations. The splittings as well as the orbital dependence of the hybridization imply that the exchange and potential couplings between magnetic impurity and conduction electrons become orbital dependent. 

When the magnetic impurity is placed in an isotropic superconductor, one thus expects five pairs of degenerate \YSR states. Similar to the $d$-levels, the \YSR states will split due to the symmetry reduction by crystal fields. One way of thinking about this splitting is as a result of the modification of the $d$-level energies and hybridizations mentioned above. Alternatively, we can first compute the Shiba states for a completely isotropic environment and then consider their splitting resulting from the symmetry reduction. 
As the results are controlled by group theory, both approaches give identical results as long as we do not attempt to compute specific values of energy levels. In the following, we briefly sketch the second approach. 

\subsection{Yu-Shiba-Rusinov states}

Let us consider a homogeneous $s$-wave superconductor whose Hamiltonian in real space can be written as
\begin{equation}
    H_{s}=\int
    dr\,\left\{\sum_{\sigma}\psi_{\sigma}^{\dagger}(\mathbf{r})\left[\frac{-\nabla^{2}}{2m}-\mu\right]\psi_{\sigma}(\mathbf{r})+\Delta^{*}\psi_{\uparrow}(\mathbf{r})\psi_{\downarrow}(\mathbf{r})+\Delta\psi_{\downarrow}^{\dagger}(\mathbf{r})\psi_{\uparrow}^{\dagger}(\mathbf{r}).\right\}
\end{equation}
Here $\psi_{\sigma}(\v{r})$ annihilates an electron with spin $\sigma$ at position $\v{r}$, $\Delta$ is the superconducting order parameter and $\mu$ is the chemical potential. 
It is convenient to represent the electron field operators in the basis of spherical waves centered at the position of the impurity atom, namely
\begin{equation}
    \psi_{\sigma}(\mathbf{r})=\sum_{klm}c_{klm\sigma}\phi_{klm}(\mathbf{r}),
\end{equation}
where 
\begin{equation}
    \phi_{klm}(\mathbf{r})=j_{l}(kr)Y_{l}^{m}(\hat{\mathbf{r}}),\quad m=-l,-l+1,\dots l
\end{equation}
with $l\in \mathbb{N}_{0}$, $j_{l}$ the spherical Bessel function of order $l$, and $Y_{l}^{m}$  the spherical harmonics of degree $l$ and order $m$. Thus, we decompose the electrons in the superconductor into different angular-momentum channels,
\begin{gather}
    H_{s}=\sum_{klm}c_{klm\sigma}^{\dagger}c_{klm\sigma}\xi_{k}+(-)^{m}\left[\Delta c_{klm\uparrow}c_{kl-m\downarrow}+\Delta
    c_{kl-m\downarrow}^{\dagger}c_{klm\uparrow}^{\dagger}\right] \\
    \xi_{k}=\frac{k^{2}}{2m}-\mu.
\end{gather}

In the situation we are considering here, the impurity atom is ${\rm Mn}^{++}$, whose ground state is $^{6}S_{5/2}$ with a half-filled $3d$ shell. It was shown \cite{Schrieffer67,Tsvelick} that, to lowest order, only the $l=2$ channel electrons get scattered due to the impurity. Hence, in the following, we will only include the $l=2$ conduction electrons in the Hamiltonian, and suppress this index. The full Hamiltonian including the impurity becomes
\begin{equation}
    H=\sum_{m=-2}^{2}\left\{ \sum_{k\sigma}c_{km\sigma}^{\dagger}c_{km\sigma}\xi_{k}+(-)^{m}\sum_{k}\left(\Delta c_{km\uparrow}c_{k-m\downarrow}+\Delta
    c_{k-m\downarrow}^{\dagger}c_{km\uparrow}^{\dagger}\right)\right\}
    +\sum_{\sigma\sigma'}\sum_{kk'}\left(
    J\mathbf{S}\cdot\gv{\sigma}_{\sigma\sigma'}+V\delta_{\sigma\sigma'}\right)c_{km\sigma}^{\dagger}c_{km\sigma'},
\end{equation}
where $J$ and $V$ are the strengths of the exchange and the potential coupling between the impurity atom and the conduction
electrons. 
If we assume the impurity spin $\v{S}$ to be aligned along $z$ 
direction, the last term in the Hamiltonian can be written as   
\[JS(c_{km\uparrow}^{\dagger}c_{km\uparrow}-c_{km\downarrow}^{\dagger}c_{km\downarrow}).\]
Introducing Nambu spinor $C_{km}=(c_{km\uparrow},c_{k-m\downarrow}^{\dagger})^{T}$, we have the Bogoliubov--de Gennes Hamiltonian
\begin{gather}
  H=\sum_{m=-2}^{2}\left\{ \sum_{k}C_{km}^{\dagger}\mathcal{H}_{s}C_{km}+\sum_{kk'}\left(JS+V\tau_{z}\right)C_{km}^{\dagger}C_{k'm}\right\} \\
    \mathcal{H}_{s}=\xi_{k}\tau_{z}+(-)^{m}\Delta\tau_{x}.
\end{gather}
Here, $\tau_\alpha$ denotes Pauli matrices in particle-hole space. 

The Green function corresponding to the above Hamiltonian fulfills the Dyson equation
\begin{equation}
    G_{kk'm}(E)=g_{km}(E)\delta_{kk'}+g_{km}(E)\left( JS+V\tau_z \right)\sum_{k_{1}}G_{k_{1}k'm}(E),
\end{equation}
where $g_{km}$ is the Green function of the homogeneous superconductor without the impurity,
\begin{equation}
    g_{km}(E)=(E-\xi_{k}\tau_{z}-(-)^{m}\Delta\tau_{x})^{-1}=\frac{E+\xi_{k}\tau_{z}+(-)^{m}\Delta\tau_{x}}{E^{2}-\xi_{k}^{2}-\Delta^{2}}.
\end{equation}
In particular, we have
\begin{equation}
\sum_{k}G_{kk'm}(E)=g_{k'm}(E)+\left[\sum_{k}g_{km}(E)\right]\left( JS+V\tau_z \right)\left[\sum_{k_{1}}G_{k_{1}k'm}(E)\right],
\end{equation}
which gives
\begin{equation}
G_{kk'm}(E)=g_{km}(E)\delta_{kk'}+(JS+V\tau_z)g_{km}(E)\left[1-\sum_{k}g_{km}(E)(JS+V\tau_z)\right]^{-1}g_{k'm}(E).
\end{equation}
One can identify the T matrix as
\begin{equation}
    T(E)=(JS+V\tau_z)\left[1-\sum_{k}g_{km}(E)\left( JS+V \tau_z \right)\right]^{-1}.
\end{equation}
Since
\begin{equation}
    \sum_{k}g_{km}(E)\simeq\int d\xi\,\nu_{0}\frac{E+(-)^{m}\Delta\tau_{x}}{E^{2}-\xi^{2}-\Delta^{2}}=\frac{-\pi\nu_{0}(E+(-)^{m}\Delta\tau_{x})}{\sqrt{\Delta^{2}-E^{2}}},
\end{equation}
with $\nu_{0}$ a one-channel density of states at the Fermi level ($\propto 1/(\pi v_F)$) for the conduction electrons,
we find that 
\begin{equation}
  T(E)=\frac{1}{\pi \nu_0}\frac{(\alpha^2 - \beta^2)E +(\alpha+\beta\tau_z)\sqrt{\Delta^2 -E^2 } + (-)^m (\alpha^2
-\beta^2 )\Delta\tau_x }{(1-\alpha^{2}+\beta^2)\sqrt{\Delta^{2}-E^{2}}+2\alpha E},
\end{equation}
whose poles give the Shiba state energies:
\begin{equation}
  E_{m}=-\Delta\frac{1-\alpha^{2}+\beta^2}{\sqrt{(1-\alpha^2+\beta)^2 + 4\alpha^2}},
\end{equation}
where $\alpha=JS\pi\nu_{0}>0$, $\beta=V\pi\nu_0$. [$J$ as defined here differs from the $J$ in the real-space representation (for s-wave scatterers) $J\psi^{\dagger}(0)\gv{\sigma}\psi(0)\cdot\v{S}$ by a normalization factor; however, the value of the dimensionless quantity $\alpha$ remains unaffected.]
This expression is the same as the one for the Shiba states induced by an exchange potential of the form of a $\delta$-function. The difference is that the Shiba states obtained here are fivefold degenerate.

\begin{table}[t]
\centering{}\protect\caption{Character table for the irreducible representations for group $C_{4v}$ and reducible
representation $D^{+}$\label{tab:Pb_100}}
\begin{tabular}{c|rrrrrcc}
 & $E$ & $2C_{4}$ & $C_{2}$ & $2\sigma_{v}$ & $2\sigma_{d}$ & linear, rotations & quadratic\tabularnewline
\hline 
$A_{1}$ & $1$ & $1$ & $1$ & $1$ & $1$ & $z$ & $x^{2}+y^{2}$, $z^{2}$\tabularnewline
$A_{2}$ & $1$ & $1$ & $1$ & $-1$ & $-1$ & $R_{z}$ & \tabularnewline
$B_{1}$ & $1$ & $-1$ & $1$ & $1$ & $-1$ &  & $x^{2}-y^{2}$\tabularnewline
$B_{2}$ & $1$ & $-1$ & $1$ & $-1$ & $1$ &  & $xy$\tabularnewline
$E$ & $2$ & $0$ & $-2$ & $0$ & $0$ & $(x,y)$ $(R_{x},R_{y})$ & $(xz,yz)$\tabularnewline
\hline 
$D^{+}$ & $5$ & $-1$ & $1$ & $1$ & $1$ &  & \tabularnewline
\hline 
\end{tabular}
\end{table}

\subsection{Crystal field splitting}

Above, the Shiba states were obtained from scattering electrons
of an isotropic superconductor off an impurity potential with a certain angular momentum
component with $l=2$. Thus, the Shiba states are 5-fold degenerate, and their wave
functions resemble the shape of $d$ atomic orbitals. The degeneracy is (partially)
removed by the crystal field describing the local environment of the magnetic impurity. The 
nature of the splitting is essentially determined by symmetry.  We briefly summarize the standard 
results of group theory which govern these splittings for the surfaces of interest in the main text. 

The point group symmetries for the Pb(001) and Pb(111) surfaces are $C_{4v}$ and $C_{3v}$ with the corresponding character tables in Tables  \ref{tab:Pb_100} and \ref{tab:Pb_111}, respectively \cite{dresselhaus}. As long as the adsorption sites respect this symmetry, we can then read off the generic multiplicities of the Shiba states. In our experiments, we find this to be the case for the Pb(001) surface as well as the for Mn$^{\rm up}_{{\rm Pb}(111)}$ site on the Pb(111) surface. 
If the adsorption site further reduces the symmetry, the Shiba states will split even further. In our experiments, we conclude that this is the case for the Mn$^{\rm down}_{{\rm Pb}(111)}$ adsorption site. 

For the case that the absorption sites respect the symmetry of the surface, we thus find from the character tables:

\begin{table}[t]
\centering{}\protect\caption{Character table for the irreducible representations for group $C_{3v}$ and reducible representation $D^{+}$\label{tab:Pb_111}}
\begin{tabular}{c|rrrcc}
 & $E$ & $2C_{3}$ & 3$\sigma_{v}$ & linear, rotations & quadratic\tabularnewline
\hline 
$A_{1}$ & $1$ & $1$ & $1$ & $z$ & $x^{2}+y^{2}$, $z^{2}$\tabularnewline
$A_{2}$ & $1$ & $1$ & $-1$ & $R_{z}$ & \tabularnewline
$E$ & $2$ & $-1$ & $0$ & $(x,y)$ $(R_{x},R_{y})$ & $(x^{2}-y^{2},xy)$ $(xz,yz)$\tabularnewline
\hline 
$D^{+}$ & $5$ & -1 & $1$ &  & \tabularnewline
\hline 
\end{tabular}
\end{table}

\begin{itemize}
\item Pb(001): 
\begin{equation}
D^{+}=A_{1}\oplus B_{1}\oplus B_{2}\oplus E.
\end{equation}
$d_{xz}$ and $d_{yz}$ orbitals are doubly degenerate, and $d_{xy}$ , $d_{x^{2}-y^{2}}$ and $d_{z^{2}}$ are nondegenerate. 

\item Pb(111):
\begin{equation}
D^{+}=A_{1}\oplus2E.
\end{equation}
$d_{x^{2}-y^{2}}$ and $d_{xy}$ are degenerate. $d_{xz}$ and $d_{yz}$ are degenerate.
$d_{z^{2}}$ is non degenerate. 
\end{itemize}

\section{Experimental Data}

\subsection{\didv maps with full contrast and at negative bias voltages}

\begin{figure}[tbh]
	\includegraphics[width=\textwidth]{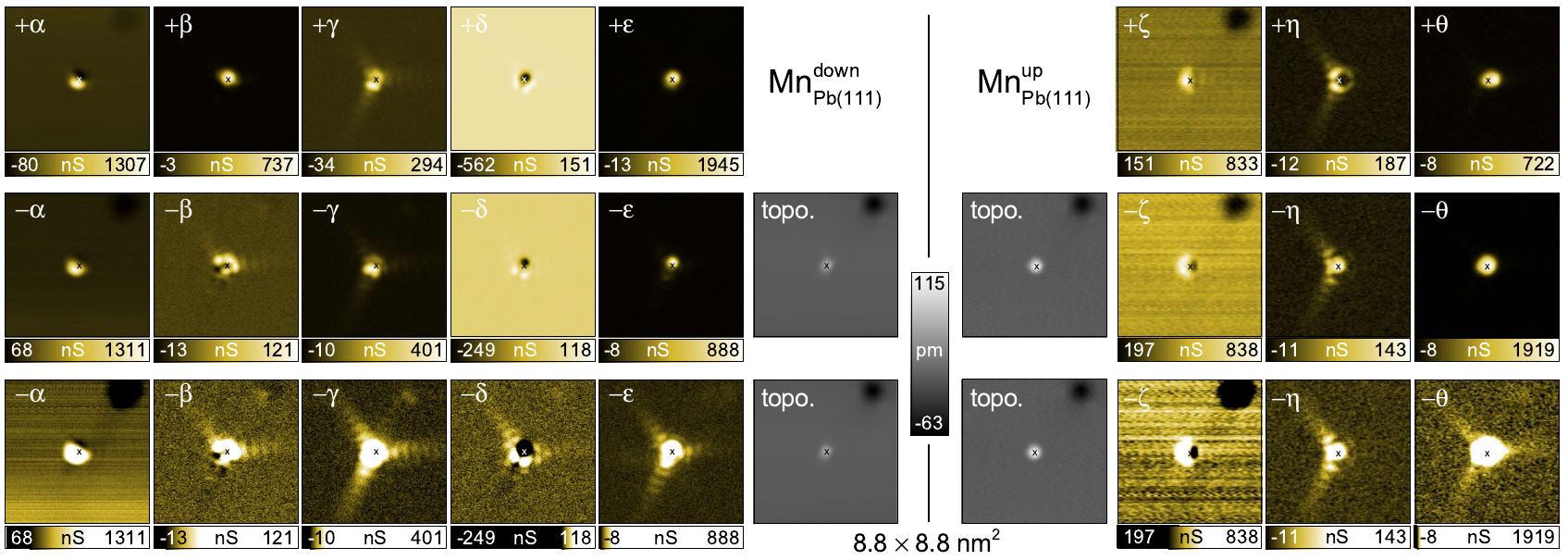}
	\caption{\didv maps of a Mn adatom on Pb(111) in the two adsorption sites denoted by \Mn{111}{down} ($\pm\alpha,\,\pm\beta,\,\pm\gamma,\,\pm\delta,\,\pm\epsilon$) and \Mn{111}{up} ($\pm\zeta,\,\pm\eta\, \pm\theta$), respectively.
	The corresponding topographies are shown. $\times$ denotes the same position in all maps. The \didv maps are recorded with the tip-sample distance adjusted in each pixel to a setpoint of $400\upA$ at $5\umV$. Lock-in modulation: $20\umuV_\mathrm{rms}$. The maps of $+\alpha$ to $+\epsilon$ and $+\zeta$ to $+\theta$
reproduce the same data as in Fig.~3 (e,f) of the main text, but with a linear color scale. The maps of the negative energy resonances $-\alpha$ to $-\epsilon$ and $-\zeta$ to $-\theta$ are shown with a linear (top) and with a stretched (bottom) color scale, respectively. Note that the dark spot in the top right corner of the imaged area is a subsurface neon inclusion \cite{Ruby14}.}
	\label{fig:1}
\end{figure}

In Fig.~3 of the main text we provided \didv maps at positive bias voltages for a Mn adatom in the \Mn{111}{down} and in the \Mn{111}{up} adsorption site. The contrast of some of the maps was stretched to emphasize the long-range patterns of the \YSR states. For completeness, we provide the same maps with a linear color scale in \Fig{1} ($+\alpha$ to $+\epsilon$ and $+\zeta$ to $+\theta$). 
We provide also \didv maps of the \YSR resonances at negative bias voltages, which show patterns similar to those at positive voltages.

\subsection{Further arguments for the orbital assignment}

\begin{figure}[tbh]
	\includegraphics[width=0.5\textwidth]{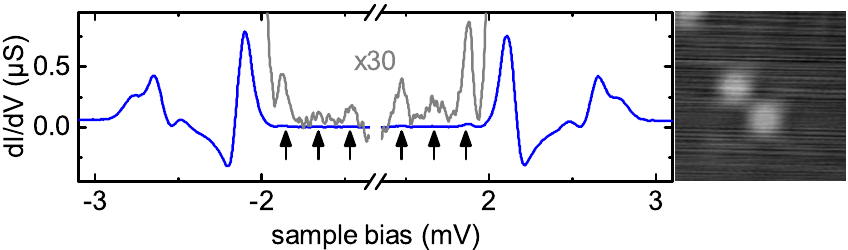}
	\caption{\didv spectrum recorded at a pair of adatoms, which lie at close distance. Splitting of the states reveals three resonances close to the original energy of $\pm\gamma$ (marked by arrows). Setpoint: $4\umV$ at $200\upA$. Lock-in modulation: $15\umuV_\mathrm{rms}$.}
	\label{fig:5}
\end{figure}

In the main text we deduced the symmetry of the scattering potential from the spatial pattern of the \YSR-states observed in the \didv maps of a Mn adatom on Pb(001). We assigned the distinct states $\alpha$, $\beta$ and $\gamma$ to originate from scattering of the Mn adatom's $d$-orbitals. The assignment of resonances $\pm\beta$ to \dorb{z^2} was unambiguous because the intensity is strongest at the center of the adatom and only weak into the \udir{110} directions. Resonances $\alpha$ and $\gamma$ originate either from scattering at \dorb{x^2-y^2} and/or from the orbitals \dorb{xz,yz} and \dorb{xy}. An assignment from the spatial shape of the \YSR state alone is ambiguous. At higher coverage, we also observe pairs of adatoms at close distance. The interaction leads to a splitting of the resonances $\pm\gamma$ into three pairs of resonances [see \Fig{5}]. This requires $\pm\gamma$ to actually consist of (at least) two resonances. 
Thus, we assigned $\pm\gamma$ resonances to scattering at the orbitals \dorb{xz,yz} and \dorb{xy}, which are degenerate in the single atom. Resonances $\pm\alpha$ then originate from scattering at \dorb{x^2-y^2}.

\subsection{Lateral decay of \didv intensity}

\begin{figure}[tbh]
	\includegraphics[width=\textwidth]{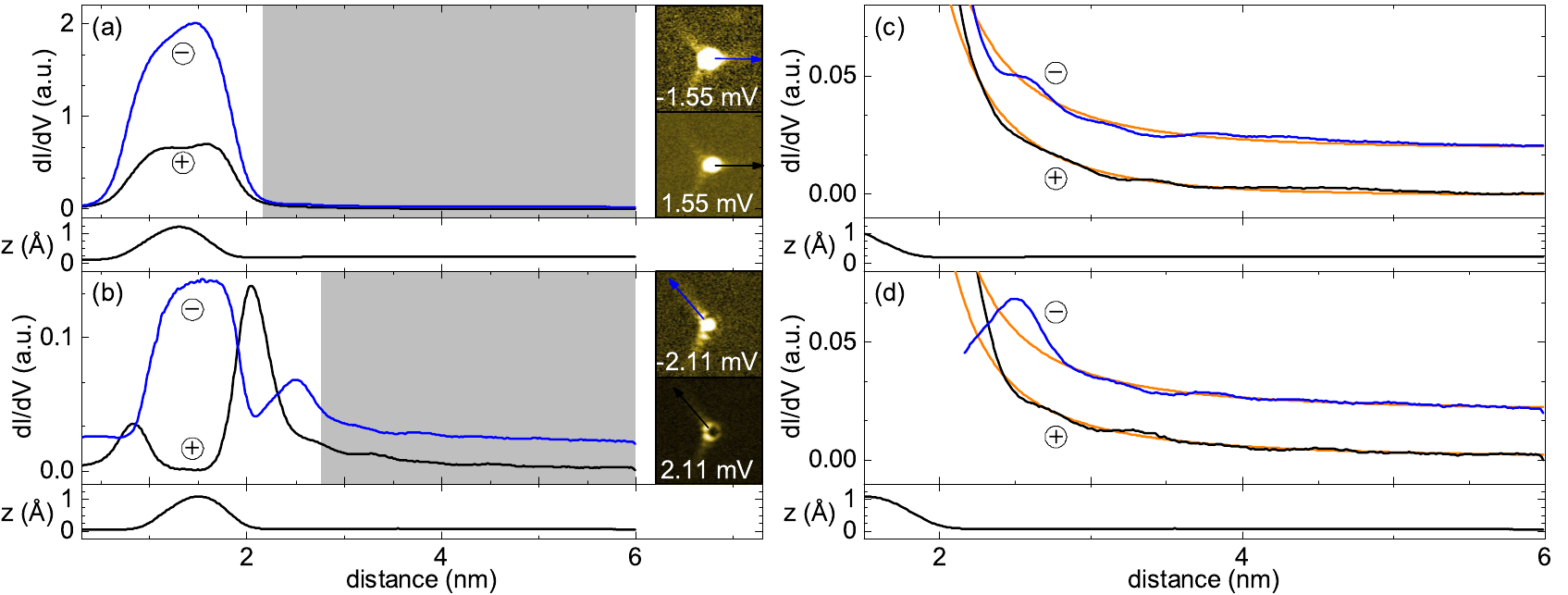}
	\caption{Lateral decay of the spectral intensity at positive (black) and negative bias (blue) of the two \YSR resonances with lowest binding energy for an adatom on Pb(111) in the \Mn{111}{up} adsorption site. Setpoint: $400\upA, 4\umV$. Lock-in modulation: $20\umuV$. The blue curves in (c,d) are offset for clarity by 0.02. (a,b) show the full profiles as extracted from high-resolution \didv maps along the \udir{110}  crystal directions. (c,d) show a zoom of the gray shaded areas in (a,b). Removal of the strongly decaying background, which is shown as orange line, leads to Fig.~3 of the main text. The $z$ profiles show the apparent height as a function of distance from the impurity center.}
	\label{fig:2}
\end{figure}

In the main text we showed the lateral decay of spectral intensity of \YSR states along the \udir{110}  high-symmetry directions in the vicinity of a \Mn{111}{up} adatom. The curves were extracted from high-resolution \didv maps at the energies of the \YSR states. To emphasize the oscillatory intensity variations, we subtracted a background $b(r)$, which is derived from the $1/r$ dependence of the \YSR wavefunction $\psi(r)$ at distances $r<\xi$ ($\xi$ is the coherence length of the superconductor):

\begin{equation}
b(r) =  \left|y_0 + \frac{1}{k \left|r - r_0\right|}\right|^2 ~.
\label{eq:Decay}
\end{equation}

Here, $k$ and $y_0$ are independent fit parameters, and $r_0$ is set to the center of the adatom. \Figure{2} shows the full datasets of the spectral intensity at positive and negative bias in (a,b). The region of interest is shaded with a gray background, and displayed in (c,d). Subtracting the decay function (orange) leads to Fig.~3 of the main text.

%
%

%
\end{document}